\documentclass{article}
\usepackage{arxiv}

\usepackage{amssymb}
\usepackage{amsmath}
\usepackage{amscd}
\usepackage{dsfont}
\usepackage{graphicx}
\usepackage{float}
\usepackage{graphics}
\usepackage[noadjust]{cite}
\usepackage{amsthm}
\usepackage{caption}
\usepackage{multirow}
\usepackage{array}
\usepackage{natbib}
\usepackage[norelsize,boxruled]{algorithm2e}
\usepackage{hyperref}
\usepackage{color}
\usepackage{indentfirst}
\usepackage{optidef}
\usepackage{url}
\hypersetup{colorlinks,linkcolor={blue},citecolor={blue},urlcolor={blue}}

\newcommand{\mb}{\mathbf}

\DeclareMathOperator*{\argmin}{arg\,min}

\renewcommand{\u}[0]{\bold{u}}                        

\renewcommand{\b}[0]{\bold{b}}                        
\renewcommand{\d}[0]{\bold{d}}                        
\renewcommand{\P}[0]{\bold{P}}                        
\renewcommand{\next}[1]{#1_{k}}                         
\newcommand{\prev}[1]{#1_{k-1}}                       

\newcommand{\m}[0]{\bold{m}}                        
\newcommand{\A}[0]{\bold{A}}                        
\newcommand{\reg}[0]{\mathcal{R}}                   
\newcommand{\I}[0]{\bold{I}}                   
\begin{document}
\title{Anderson accelerated augmented Lagrangian for extended waveform inversion}

\author{\href{https://orcid.org/0000-0002-2080-9697}{\includegraphics[scale=0.06]{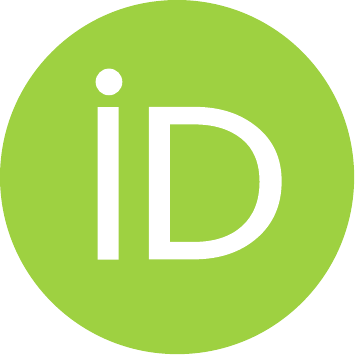}\hspace{1mm}Kamal Aghazade} \\
  Institute of Geophysics, University of Tehran, Tehran, Iran. 
  \texttt{aghazade.kamal@ut.ac.ir}
\And
  \href{https://orcid.org/0000-0002-9879-2944}{\includegraphics[scale=0.06]{orcid.pdf}\hspace{1mm}Ali Gholami} \\
  Institute of Geophysics, University of Tehran, Tehran, Iran.
  \texttt{agholami@ut.ac.ir} \\ 
  \And
  \href{http://orcid.org/0000-0003-1805-1132}{\includegraphics[scale=0.06]{orcid.pdf}\hspace{1mm}Hossein S. Aghamiry} \\
  University Cote d'Azur - CNRS - IRD - OCA, Geoazur, Valbonne, France. 
  \texttt{aghamiry@geoazur.unice.fr} 
  \And
\href{http://orcid.org/0000-0002-4981-4967}{\includegraphics[scale=0.06]{orcid.pdf}\hspace{1mm}St\'ephane Operto} \\ 
  University Cote d'Azur - CNRS - IRD - OCA, Geoazur, Valbonne, France. 
  \texttt{operto@geoazur.unice.fr}
  }

\renewcommand{\shorttitle}{Anderson Accelerated IR-WRI, Aghazade et al.}

\maketitle

\begin{abstract}
The augmented Lagrangian (AL) method provides a flexible and efficient framework for solving extended-space full-waveform inversion (FWI), a constrained nonlinear optimization problem whereby we seek model parameters and wavefields that minimize the data residuals and satisfy the wave equation constraint. The AL-based wavefield reconstruction inversion, also known as iteratively refined wavefield reconstruction inversion, extends the search space of FWI in the source dimension and decreases sensitivity of the inversion to the initial model accuracy. 
Furthermore, it benefits from the advantages of the alternating direction method of multipliers (ADMM), such as generality and decomposability for dealing with non-differentiable regularizers, e.g., total variation regularization, and large scale problems, respectively. 
In practice any extension of the method aiming at improving its convergence and decreasing the number of wave-equation solves would have a great importance. To achieve this goal, we recast the method as a general fixed-point iteration problem, which enables us to apply sophisticated acceleration strategies like Anderson acceleration. 
The accelerated algorithm stores a predefined number of previous iterates and uses their linear combination together with the current iteration to predict the next iteration. 
We investigate the performance of the proposed accelerated algorithm on a simple checkerboard model and the benchmark Marmousi II and 2004 BP salt models through numerical examples. These numerical results confirm the effectiveness of the proposed algorithm in terms of convergence rate and the quality of the final estimated model.
\end{abstract}

\section{INTRODUCTION}
Full waveform inversion (FWI) is the state-of-the-art inversion strategy for delineating subsurface physical properties. It is a nonlinear data matching problem that utilizes the entire content of recorded data to extract subsurface parameters at the wavelength scale resolution \citep{tarantola1984inversion, virieux2009overview,fichtner2011resolution}.  

From the mathematical point of view, FWI is a nonlinearly constrained optimization problem, in which a suitable regularization term is minimized subject to satisfying the wave-equation and the observation equation constraints \citep{haber2000optimization}. The former constraint requires the model parameters to be consistent with physics and the latter requires consistency with the observed data.
Traditional FWI algorithms use the variable projection method to eliminate the wavefields from the equations \citep[e.g., ][] {pratt1998gauss,virieux2009overview,operto2006crustal,brossier2010data}. 
In these methods, the wave-equation constraint is solved exactly at each iteration, leading to a reduction in the search space of the optimization problem. 
The resulting objective function is traditionally solved by gradient-based local optimization algorithms such as the preconditioned steepest descent and the nonlinear conjugate gradient methods. Recently, the l-BFGS quasi-Newton and truncated Newton methods have been proposed to solve FWI for a faster convergence rate by including second-order information into the inversion \citep{metivier2013full}. The reader is referred to \citet{Metivier_2015_SOT} for an overview of classical optimization algorithms and their computer implementation.

The main issue with the reduced-space approach is the sensitivity of the inversion to the accuracy of the initial model. 
The performance is being limited to kinematically accurate starting models or the availability of low-frequency content in data. Otherwise, the method may converge to a local minimum \citep{virieux2009overview}.
Several attempts have been made to increase the robustness to the initial model by modifying the misfit function such as those based on correlation or deconvolution \citep{van2010correlation,Luo_2011_DBO}, adaptive matching filters \citep{Warner_2016_AWI}, and optimal transport distance \citep{Yang_2017_AOT,Metivier_2019_GOT}. All of these methods are implemented with exact satisfaction of the wave equation.

Some approaches such as the contrast-source method  \citep{Abubakar_2009_FDC} and the wavefield reconstruction inversion method \citep{van2013mitigating} solve the original nonlinearly constrained optimization problem with a penalty method to implement the wave equation as a soft constraint \citep{Abubakar_2009_FDC,van2013mitigating}. In these approaches, a quadratic penalty term  corresponding to the wave-equation misfit function is added to the data-misfit function where a constant penalty parameter balances the relative weight of the two misfit functions. The wave-equation relaxation extends the search space by considering the wavefield as an unknown variable in addition to the model parameters.
This extension allows for the data to be closely matched with inaccurate subsurface models from the early FWI iterations, hence increasing the robustness of the method to the initial solution. This improved data fit is achieved by solving an augmented wave equation with the data or observation equation, leading to the so-called data-assimilated wavefields \citep{Aghamiry_2020_AED}. The main issue of penalty methods is however to provide  approximate solution of the wave equation at the convergence point when a fixed penalty parameter is used, unless this value is dynamically increased via a continuation strategy \citep{Fu_2017_DPM}.

In order to overcome this issue, \citet{aghamiry2019improving} proposed the iteratively refined wavefield reconstruction inversion (IR-WRI) method that solves the original nonlinear constrained optimization problem with the augmented Lagrangian (AL) method. The AL combines the penalty objective function and a Lagrangian function, the latter providing a second leverage that allows for the wavefield solution to satisfy the wave equation accurately at the convergence point with a fixed penalty parameter \citep{nocedal2006numerical}. 
Originally, IR-WRI was implemented in the frequency-space domain where the relaxed wave equation (namely, the augmented wave equation with the data or observation equation) can be solved more easily. Recently, the method has been also formulated in the data space for efficient time-domain implementation \citep{gholami2020extended,Gholami_2021_CSI,Gholami_2021_ADR}.  

The IR-WRI is a biaffine problem and benefits from the advantages of the alternating direction method of multipliers (ADMM) \citep{boyd2011distributed} such as generality and decomposability. The former property means that the algorithm can deal with both differentiable and non-differentiable regularizers. The latter property makes the algorithm suitable for dealing with large-scale problems because it allows one to break down the optimization task into a set of smaller tasks. 
The IR-WRI decomposes the problem into two subproblems (i.e., data-assimilated wavefield reconstruction and parameter estimation) that are solved in an alternating fashion during iterations, the most computationally-expensive task being the former. The main drawback of the method is its slow convergence, which can limit its applicability in 3D field case studies. 

This limitation prompts us to investigate acceleration methods for improving the convergence rate of IR-WRI, i.e. decrease the number of iterations required for reaching a satisfactory solution and thus decrease the computational burden.
Inspired by \citet{zhang2019accelerating}, we recast the IR-WRI iterations as a general nonlinear bivariate fixed-point problem for the primal variables and dual variables (Lagrange multipliers). This formulation allows us to investigate acceleration strategies that are suitable to solve fixed-point problems. 
Among the accelerating strategies for fixed point iterations, the Anderson Acceleration (AA) \citep{anderson1965iterative} has gained considerable interest in the optimization community \citep{walker2011anderson}. The AA strategy is based on storing a (pre-defined) history of previous iterations and predicts the new iteration using a weighted linear combination of the available history. 
AA shares the same characteristics as the quasi-Newton method for accelerating the fixed point iteration 
\citep{scieur2019generalized}. 
Recently, \citet{yang2020anderson} has compared limited memory-BFGS (l-BFGS) with AA for classical FWI and reverse time migration (RTM) and showed that the AA method outperforms l-BFGS in terms of convergence rate.

This paper investigates the application of AA for the IR-WRI within the following structure: a brief review of the AA method for acceleration of fixed point iteration is proposed, followed by a mathematical description of the IR-WRI method. Then the proposed accelerated IR-WRI is analyzed. The performance of the proposed method is assessed in the numerical examples section using a simple checkerboard model and the benchmark Marmousi II and 2004 BP salt models. Finally, some conclusions are provided.
\section{METHOD}

\subsection{Fixed point iteration and Anderson acceleration}
\begin{figure}[!htb]
 	\centering
 	\includegraphics[trim={0.4cm 0.4cm 0.4cm 0.2cm},clip,width=.45\textwidth]{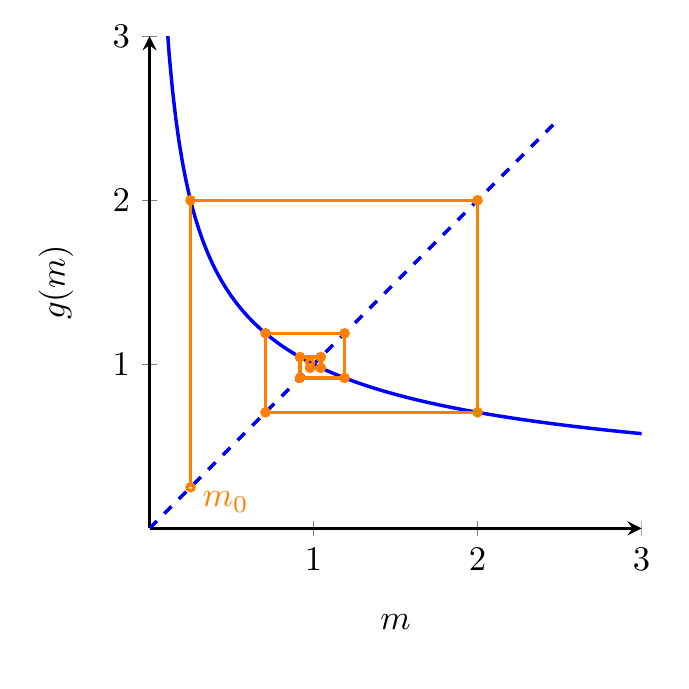}
 	\caption{Schematic representation of fixed point iterations. It starts with initial estimate $m_0$ on $y=m$, i.e. the dashed line, and then move vertically to the solid blue curve. Then it moves horizontally to dashed line and again vertically to the solid curve. This procedure continues until convergence.}
 	\label{FPI:blue}
 \end{figure}

Consider the problem of solving the nonlinear equation $f(\m)=\bold{0}$ for model parameters vector $\m$, where $f$ is a 
mapping function. Then $\m$ may be defined as a fixed point of a properly defined 
mapping $g$ \citep{richard1985burden}
\begin{equation}\label{FP:red}
\m=g(\m). 
\end{equation}
The fixed point iteration is a well-established approach for the solution of an fixed point problem $\m = g(\m)$  which is indeed equal to solving $f(\m)=\m-g(\m)=\bold{0}$. Fixed point iteration solves the problem in equation \ref{FP:red} through iterations: 
\begin{equation}\label{fiexd:red}
\next{\m} = g(\prev{\m}), \qquad k=1,2,... 
\end{equation} 
where $k$ denotes the iteration number.\\ 
Fixed point iterations solve two functions simultaneously: $\m$ and $g(\m)$. The intersection point of these two functions is the solution of $\m = g(\m)$, and thus $f(\m) = \bold{0}$. This process is illustrated in Fig. \ref{FPI:blue} for a 1D case.
Fixed point iteration arises in various fields of science \citep [see][ and references therein]{walker2011anderson}. One of the main issues related to the fixed point iteration is that the iterations may have a slow convergence (linear convergence). Therefore, accelerating the convergence rate of the fixed point iteration has attracted considerable interest.


Anderson Acceleration (AA) \citep{anderson1965iterative} is among the most popular techniques to speed up the convergence of fixed point iteration \citep{walker2011anderson,scieur2018nonlinear,bollapragada2018nonlinear}. The key idea behind the AA strategy is to maintain the history of $h$ recent iterations and predict the new iteration by using a linear combination of this history where the weights are extracted by solving an optimization problem. AA seeks to speed up the convergence of the fixed point iteration problem in equation \ref{fiexd:red} by decreasing the following residual \citep{walker2011anderson}:
\begin{equation}\label{AA1:red}
	  {f}(\m) = \m-{g}(\m).
\end{equation} 
The linear combination of $h+1$ previous iterations, i.e. $\m_{k}, \m_{k-1},...,\m_{k-h}$, under the fixed point mapping $g$ reads: 
\begin{equation}\label{AA2:red}
	\m_{k+1} = \sum_{j=0}^{h} \theta_{j} g(\m_{k-h+j}),
\end{equation} 
where coefficients $\theta_0,\theta_{1},...,\theta_{h}$ are the weights for constructing the new iteration and they are extracted by solving the following constrained optimization: 
\begin{mini} 
{\boldsymbol{\theta}}{\|\sum_{j=0}^{h} {\theta}_j f(\m_{k-h+j})\|_2^2}
{\label{eq:main}}{}
\addConstraint {\sum_{j=0}^{h} {\theta}_{j} }{=1},
\end{mini}
where $\|\cdot\|_2$ is the $\ell_2$-norm. 
The acceleration is nonlinear because the coefficients are updated at each iteration. 
These coefficients are calculated very simply as \citep{walker2011anderson}
\begin{equation} \label{theta}
\begin{cases}
\theta_0 = \gamma_0 \\
\theta_j = \gamma_j - \gamma_{j-1}~\text{for}~j=1,...,h-1 \\
\theta_h = 1 - \gamma_{h-1},
\end{cases}
\end{equation}
where $\boldsymbol{\gamma}$ is defined as
\begin{equation}\label{get_gama:red}
\boldsymbol{\gamma} = (\bold{F}^{T}\bold{F})^{-1}\bold{F}^{T}f(\m_k),
\end{equation}
with
\begin{equation}
\bold{F} = 
\begin{bmatrix}
\delta f_{k-h} & \delta f_{k-h+1} & \cdots & \delta f_{k-1}
\end{bmatrix},
\end{equation}
and $\delta f_{j} = f(\m_{j+1})-f(\m_{j})$.

\subsection{Iteratively refined WRI}
We consider FWI as the following nonlinear PDE-constrained optimization \citep{Aghamiry_2019_CRO}: 
\begin{mini} 
{\u,\m \in \mathcal{M}}{\reg(\m)}
{\label{constr_eq:red}}{}
\addConstraint  {\A(\m)\u}{=\b} \text{~~~and~~~}
{\P\u}{=\d},
\end{mini}
where $\P$ is the sampling operator, $\u$ is the seismic wavefield, $\mb{d}$ is the observed data, $\A(\m)$ is discretized wave-equation operator, $\m \in \mathcal{M}$ is the model parameters, $\mb{b}$ is the source term, $\mathcal{R}(\bold{m})$ is an appropriate regularization function on the model space and $\mathcal{M}$ is a convex set defined according to our prior knowledge of $\bold{m}$. For example, if we know the lower bound, $\bold{m}_{min}$, and upper bound, $\bold{m}_{max}$, of $\bold{m}$ then
\begin{equation}
\mathcal{M} = \{\bold{m} \vert \bold{m}_{min} \leq \bold{m} \leq \bold{m}_{max}\}.
\end{equation} 
For the sake of compactness, we present the formulation for a single source and single-frequency pair. The extension to the multi-source and multi-frequency is straightforward and can be achieved by summation over sources and frequencies. 
The AL method replaces the constrained optimization described in equation \ref{constr_eq:red} with the following minimax optimization \citep{aghamiry2019improving}:
\begin{equation}\label{AL:red}
\min_{\mb{u},\mb{m}\in \mathcal{M}}\max_{\boldsymbol{\lambda},\boldsymbol{\nu}} ~~\mathcal{L}(\mb{u},\mb{m},\boldsymbol{\lambda},\boldsymbol{\nu}),
\end{equation}
where
\begin{equation}\label{AL:fun}
\mathcal{L}(\mb{u},\mb{m},\boldsymbol{\lambda},\boldsymbol{\nu})=
\reg(\m)+ \frac{\alpha}{2} \|\mb{A}(\mb{m})\mb{u}-\mb{b}\|_{2}^{2} +\frac{\beta}{2}\|\mb{P}\mb{u}-\mb{d}\|_{2}^{2}- \boldsymbol{\lambda}^{T} [\mb{A}(\mb{m})\mb{u}-\mb{b}]- \boldsymbol{\nu}^{T} [\mb{P}\mb{u}-\mb{d}],
\end{equation}
 $\alpha$ and $\beta$ are penalty parameters, and $\boldsymbol{\lambda}$ and $\boldsymbol{\nu}$ are the vectors of Lagrange multipliers or dual variables. 
The first three terms of this objective function is the penalty formulation of the equation \ref{constr_eq:red} and the rest of them are the Lagrangian terms, which force the wave equation and observation equation to be satisfied at the convergence point even for a finite value of $\alpha$ and $\beta$. This is an advantage of the AL method over the penalty method.
Beginning with an initial model $\bold{m}_{0}$, the IR-WRI solves this minimax problem, equation \ref{AL:red}, via the following iteration \citep{aghamiry2019improving}:
\begin{subequations}
\label{admmm_scaled:red}
\begin{align}
		 \mb{u}_{k+1}  =&  \argmin_\mathbf{u}  \mathcal{L}(\mb{u},\mb{m}_k,\boldsymbol{\lambda}_k,\boldsymbol{\nu}_k), \label{sadmm_u}  \\
		 \mb{m}_{k+1}  =&  \argmin_\mb{m\in \mathcal{M}} ~~\mathcal{L}(\mb{u}_{k+1},\mb{m},\boldsymbol{\lambda}_k,\boldsymbol{\nu}_k), \label{sadmm_m} \\
		 \boldsymbol{\lambda}_{k+1}  =&  \boldsymbol{\lambda}_k - \alpha(\mb{A}(\mb{m}_{k+1})\mb{u}_{k+1} - \mb{b}) \label{sadmm_b},\\
		  \boldsymbol{\nu}_{k+1}  =&  \boldsymbol{\nu}_{k}- \beta(\mb{P}\mb{u}_{k+1} - \mb{d}), \label{sadmm_d}
\end{align}
\end{subequations}
where the Lagrange multipliers are updated through a gradient ascent scheme in equations~\ref{sadmm_d}-\ref{sadmm_b} to partially maximize the AL function.
These Lagrange multipliers are formed by the running sum of the constraint residuals in iterations and are re-injected in the misfit functions to iteratively refine the variables of the bi-convex optimization at each iteration. The defect correction performed by the Lagrange multipliers is effective for improving the convergence of the algorithm and is similar to that used in the Bregman iterative regularization \citep{Osher_2005_IRM}.
We refer the reader to \citet{Aghamiry_2019_CRO, Aghamiry_2020_FWI} for the closed-form expression of the optimization subproblems \ref{sadmm_u} and \ref{sadmm_m} with bound constraints and different regularizations, as well as some details about the tuning of the penalty parameters. 

\section{IR-WRI as a fixed point iteration}
From equation \ref{sadmm_u}, the wavefield at iteration $k+1$ is a function of the primal-dual triplet ($\mb{m}_k,\boldsymbol{\lambda}_k,\boldsymbol{\nu}_k$). The model $\m_{k+1}$, equation \ref{sadmm_m}, is a function of the triplet ($\u_{k+1},\boldsymbol{\lambda}_k,\boldsymbol{\nu}_k$) and hence is a function of the primal-dual triplet ($\mb{m}_k,\boldsymbol{\lambda}_k,\boldsymbol{\nu}_k$). Furthermore, $\boldsymbol{\lambda}_{k+1}$, equation \ref{sadmm_b}, is a function of ($\u_{k+1},\m_{k+1},\boldsymbol{\lambda}_{k}$) and hence is also a function of the primal-dual triplet ($\mb{m}_k,\boldsymbol{\lambda}_k,\boldsymbol{\nu}_k$). Finally, $\boldsymbol{\nu}_{k+1}$ is a function of $\u_{k+1}$ and $\boldsymbol{\nu}_k$ and  
%
hence a function of the primal-dual triplet ($\mb{m}_k,\boldsymbol{\lambda}_k,\boldsymbol{\nu}_k$).
Consequently, the IR-WRI iterations reads
\begin{subequations}
\label{AIR-WRI:red}
	\begin{align}	 
		 \mathbf{m}_{k+1}  =&  g_m(\mb{m}_k,\boldsymbol{\lambda}_k,\boldsymbol{\nu}_k)= \argmin_\mb{m\in \mathcal{M}} ~~\mathcal{L}(\u(\mb{m}_k,\boldsymbol{\lambda}_k,\boldsymbol{\nu}_k),\mb{m},\boldsymbol{\lambda}_k,\boldsymbol{\nu}_k), \label{AIR-WRI:red_a}\\
\boldsymbol{\lambda}_{k+1}  =&  g_{{\lambda}}(\mb{m}_k,\boldsymbol{\lambda}_k,\boldsymbol{\nu}_k)= 
\boldsymbol{\lambda}_k - \alpha(\mb{A}(g_m(\mb{m}_k,\boldsymbol{\lambda}_k,\boldsymbol{\nu}_k))\u(\mb{m}_k,\boldsymbol{\lambda}_k,\boldsymbol{\nu}_k) - \mb{b}), \label{AIR-WRI:red_b}\\
\boldsymbol{\nu}_{k+1}  =&  g_{\nu}(\mb{m}_k,\boldsymbol{\lambda}_k,\boldsymbol{\nu}_k)= \boldsymbol{\nu}_{k}-\beta(\mathbf{P}\u(\mb{m}_k,\boldsymbol{\lambda}_k,\boldsymbol{\nu}_k)-\mathbf{d}), \label{AIR-WRI:red_c}
	\end{align}
\end{subequations} 
where $\u(\mb{m}_k,\boldsymbol{\lambda}_k,\boldsymbol{\nu}_k)$ is the solution of equation \ref{sadmm_u}. The subproblem \eqref{AIR-WRI:red_a} can be solved using variable splitting methods and ADMM when a non-differentiable regularization function are used, e.g. total variation (TV) \citep{aghamiry2019implementing}, combination of TV and Tikhonov (TT) \citep{Aghamiry_2019_CRO}, or those that are based on black box denoisers \citep{Aghamiry_2020_FWI}. The variable splitting methods break down the original problem \eqref{AIR-WRI:red_a} into some easy to solve subproblems, i.e. a least-squares problem to update $\m_{k+1}$, some proximity/denoising problems to find auxiliary primal variables and some dual variables which keep the summation of the mismatch between the auxiliary variables and $\m_{k+1}$. For more detail interested readers can refer to \citet{Goldstein_2009_SBM} and the above mentioned references. When regularization and bound constraints are also implemented, we can consider the new primal and dual variables as the input of the fixed point problem in addition to the $\m_k$, $\boldsymbol{\lambda}_k$ and $\boldsymbol{\nu}_k$. For the ease of notations, hereafter $\mathfrak{P}$ denotes a long vector consisting of all the primal variables and similarly $\mathfrak{D}$ denotes a long vector consisting of all the dual variables. 
Then we can write the IR-WRI as the following general bivariate fixed point iteration:
\begin{equation}\label{Fixed_admm:red}
(\mathfrak{P}_{k+1}, \mathfrak{D}_{k+1})  = g(\mathfrak{P}_{k}, \mathfrak{D}_{k}),
\end{equation}
where $g$ is the corresponding fixed point mapping function which maps the current iterate ($\mathfrak{P}_k,\mathfrak{D}_k$) to the next iterate ($\mathfrak{P}_{k+1},\mathfrak{D}_{k+1}$).
By recasting IR-WRI as an fixed point iteration, equation  \ref{Fixed_admm:red}, the convergence of the algorithm can be improved by employing generally accepted acceleration techniques such as AA. In this way, traditional IR-WRI can be viewed as Anderson accelerated IR-WRI with zero history. In contrast, accelerated IR-WRI captures the information from the history of \textit{h} previous iterates and reproduces the current iteration through weighted linear combinations of the history. The accelerated IR-WRI procedure in its simplest form is given in Algorithm 1. 
\SetAlgorithmName{Algorithm 1}{}{}
\begin{algorithm}
\label{Alg_AIR-WRI:blue}
	  	\SetAlgoLined
	  	Given  $\b, \d, \m_0$\\
	  	Set $k\leftarrow0$, $\boldsymbol{\lambda}_0 \leftarrow \b$, $\boldsymbol{\nu}_0 \leftarrow \d$ and the rest of primal/dual variables equal to zero \\ 
	 \While{conditions not satisfied}{
	  	Calculate the coefficients $\boldsymbol{\theta}$ using equations \ref{theta} and \ref{get_gama:red}\\
	  	 $(\mathfrak{P}_{k+1},\mathfrak{D}_{k+1}) \leftarrow \sum_{j=0}^{h} \theta_{j} g(\mathfrak{P}_{k-h+j},\mathfrak{D}_{k-h+j})$ 
	  	}
\caption{Accelerated IR-WRI} 
\end{algorithm}
\SetAlgorithmName{Algorithm 2}{}{}
\begin{algorithm}
\label{aa_safe:alg}
\SetAlgoLined
Given $\b, \d, \m_0$ \\
 	Set $k\leftarrow0$, $\boldsymbol{\lambda}_0 \leftarrow \b$, $\boldsymbol{\nu}_0 \leftarrow \d$ and the rest of dual variables equal to zero \\ 	     \While{conditions not satisfied}{
	    \vspace{.2cm}
	  	 \nlset{1}   $(\mathfrak{P}_{k+1},\mathfrak{D}_{k+1}) \leftarrow g(\mathfrak{P}_k,\mathfrak{D}_k)$ \\           \nlset{2}Calculate the coefficients $\boldsymbol{\theta}$ using equations \ref{theta} and \ref{get_gama:red}\\
	  	\nlset{3}	$(\mathfrak{P}^{AA},\mathfrak{D}^{AA}) \leftarrow \sum_{j=0}^{h} \theta_{j}g(\mathfrak{P}_{k-h+j},\mathfrak{D}_{k-h+j})$ \\
	  	\vspace{.2cm}	   
	   \nlset{4}  $(\mathfrak{P}^{\star},\mathfrak{D}^{\star}) \leftarrow g(\mathfrak{P}^{AA},\mathfrak{D}^{AA})$ \\
	  
       \nlset{5}     \If{$res(\mathfrak{P}^{\star},\mathfrak{D}^{\star})<res(\mathfrak{P}_{k+1},\mathfrak{D}_{k+1})$}{
            \vspace{.1cm}
       \nlset{6}      $ (\mathfrak{P}_{k+1},\mathfrak{D}_{k+1}) = (\mathfrak{P}^{AA},\mathfrak{D}^{AA})$  \\
             \nlset{7} $ (\mathfrak{P}_{k+2},\mathfrak{D}_{k+2}) = (\mathfrak{P}^{\star},\mathfrak{D}^{\star})$ 
                \hspace*{4.5em}%
        \rlap{\smash{$\left.\begin{array}{@{}c@{}}\\{}\\{}\\{}\\{}\\{}\\{}\end{array}\color{black}\right\}%
          \color{black}\begin{tabular}{l}Safeguarding\\ ~~~~~step\end{tabular}$}}

 \nlset{8} $k=k+2$  \\ 
 \nlset{9}    Go to line 2 \\ 
	  	}
\nlset{10}	  $k=k+1$
	  	}
 \caption{Accelerated IR-WRI with safeguarding step}
\end{algorithm}

\subsection{Damped AA}
Due to the possible ill-posedness or rank-deficiency of the matrix $\bold{F}$ in equation \ref{get_gama:red}, there is a possibility of ill-conditioning of the AA iterations.  To address this problem, previous researches proposed a damped or regularized version of the AA-algorithm \citep{zhang2018globally,henderson2019damped}, in which $\boldsymbol{\gamma} $ is computed as
\begin{equation}\label{damped:red}
\boldsymbol{\gamma} = (\bold{F}^{T}\bold{F} + \eta\I)^{-1}\bold{F}^{T}f(\m_k),
\end{equation}
where $\eta>0$ and $\I$ is the identity matrix. 
%

\subsection{Safeguarded AA}	 
Embedding a safeguarding step in the AA algorithm during iterations may be required for improving the performance. The safeguard aims for a more stable convergence of the algorithm \citep{fang2009two}. Safeguarding \citep{zhang2018globally} is a step that sets a condition (or set of conditions) by which one can guarantee the decrease of a pre-defined residual. For example, in a situation that the AA output does not decrease the residual(s), in a hybrid manner, one may replace the result of the AA with the Krasnosel'ski{\i}-Mann (or $\alpha$-averaged) iteration \citep{mann1953mean,krasnosel1955two}, which is also another broadly used method for solving fixed point iteration.
 
Algorithm 2 reviews our safeguarding strategy in AA to accelerate IR-WRI. 
Suppose that, at iteration $k$, we have access to $h$ previous iterates ($\mathfrak{P}_{k-j},\mathfrak{D}_{k-j}$) $j=0,...,h-1$, we perform a single iteration of the IR-WRI (by applying $g$ on ($\mathfrak{P}_{k},\mathfrak{D}_{k}$), line 1) and get ($\mathfrak{P}_{k+1},\mathfrak{D}_{k+1}$). 
Then the weights $\theta_j$ are computed (line 2), and the iterates are linearly combined at line 3 to get ($\mathfrak{P}^{AA},\mathfrak{D}^{AA}$). 
We need to decide whether ($\mathfrak{P}^{AA},\mathfrak{D}^{AA}$) is acceptable or not, based on the residual value (the sum of the data and source residuals). It is worth mentioning that calculating the residual requires evaluation of $g$.  Thus the mapping $g$ is applied on ($\mathfrak{P}^{AA},\mathfrak{D}^{AA}$) and the output is called ($\mathfrak{P}^*,\mathfrak{D}^*$) (line 4). If the corresponding residual is less than that of ($\mathfrak{P}_{k+1},\mathfrak{D}_{k+1}$) (line 5), then we accept ($\mathfrak{P}^{AA},\mathfrak{D}^{AA}$) as ($\mathfrak{P}_{k+1},\mathfrak{D}_{k+1}$) and  ($\mathfrak{P}^*,\mathfrak{D}^*$) as ($\mathfrak{P}_{k+2},\mathfrak{D}_{k+2}$) (lines 6 and 7) and go to line 2. Otherwise, ($\mathfrak{P}^{AA},\mathfrak{D}^{AA}$) is rejected and the algorithm continuous with the output of line 1.

\subsection{The AA history}	
The AA update is based on a predefined history of the previous iterations; thus, we need to store $h$ previous iterates (model parameters and dual variables). This is the cost we should pay for increasing the convergence rate by AA. The computational burden of implementing AA is not demanding because the weighting coefficients are obtained by inverting an $(h+1)\times (h+1)$ system, equation \ref{damped:red}. 
The choice of $h$ depends on the problem's complexity. The larger the value of $h$, the more information from the previous iterations is incorporated. However, higher value of $h$ may increase the ill-conditioning of the problem plus extra storage specifications. Additionally, update information from the previous iterations may decrease the convergence rate of AA \citep{walker2011anderson}.
For all tests in this paper, we use a small history parameter ($h\leq 10$).


\section{NUMERICAL EXAMPLES}
We assess the performance of the proposed accelerated IR-WRI strategy with a checkerboard model and 2D mono parameter synthetic benchmarks. We performed frequency-domain finite-difference modeling with a 9-point stencil and perfectly-matched layer (PML) \citep{chen2013optimal}. Also, we use a fixed penalty parameter during the inversion, though it can be increased gradually up to a pre-defined value. Additionally, we use bound constraint (BC) using the lower and upper bounds of the true model in all of the examples, that means, the WRI and IR-WRI methods are implemented with BC by default. We also implement bound-constrained TV-regularization (BTV) with and without AA in our tests to compare the performance of each and their combination with AA. 
We use the calculated model error defined as $\|\m^*-\m_{k}\|_2/\|\m^*\|_2$ to assess the convergent of the proposed methods.
%
\subsection{Checkerboard model}
\begin{figure}[!h]
   \centering
   \includegraphics[width=.65\textwidth]{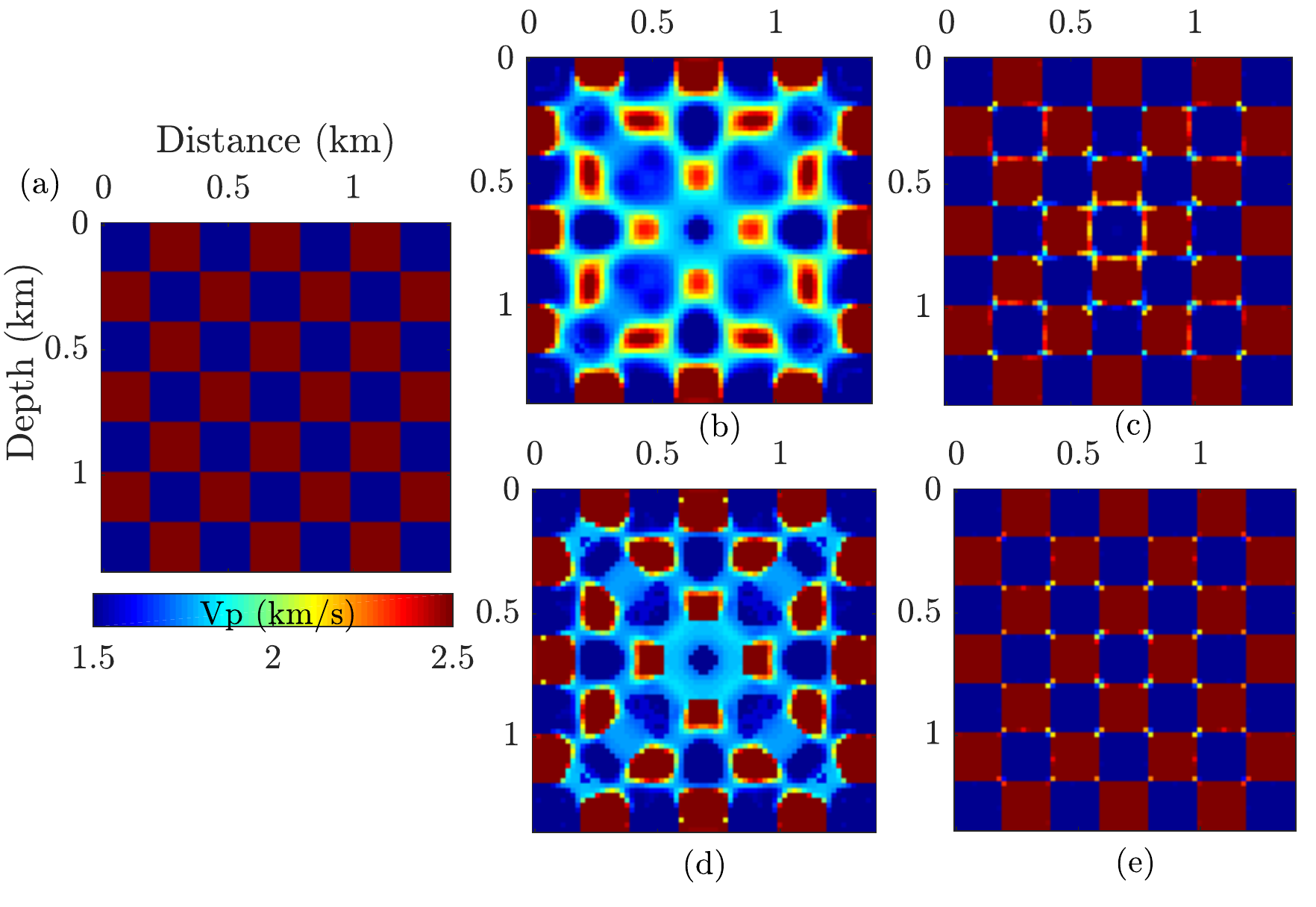}
   \caption{The checkerboard test. (a) True velocity model, (b-e) inverted velocity models by (b) WRI, (c) IR-WRI, (d) WRI with AA, and (e) IR-WRI with AA}
   \label{checker:blue}
 \end{figure}

\begin{figure}[!h]
   \centering
   \includegraphics[width=.45\textwidth]{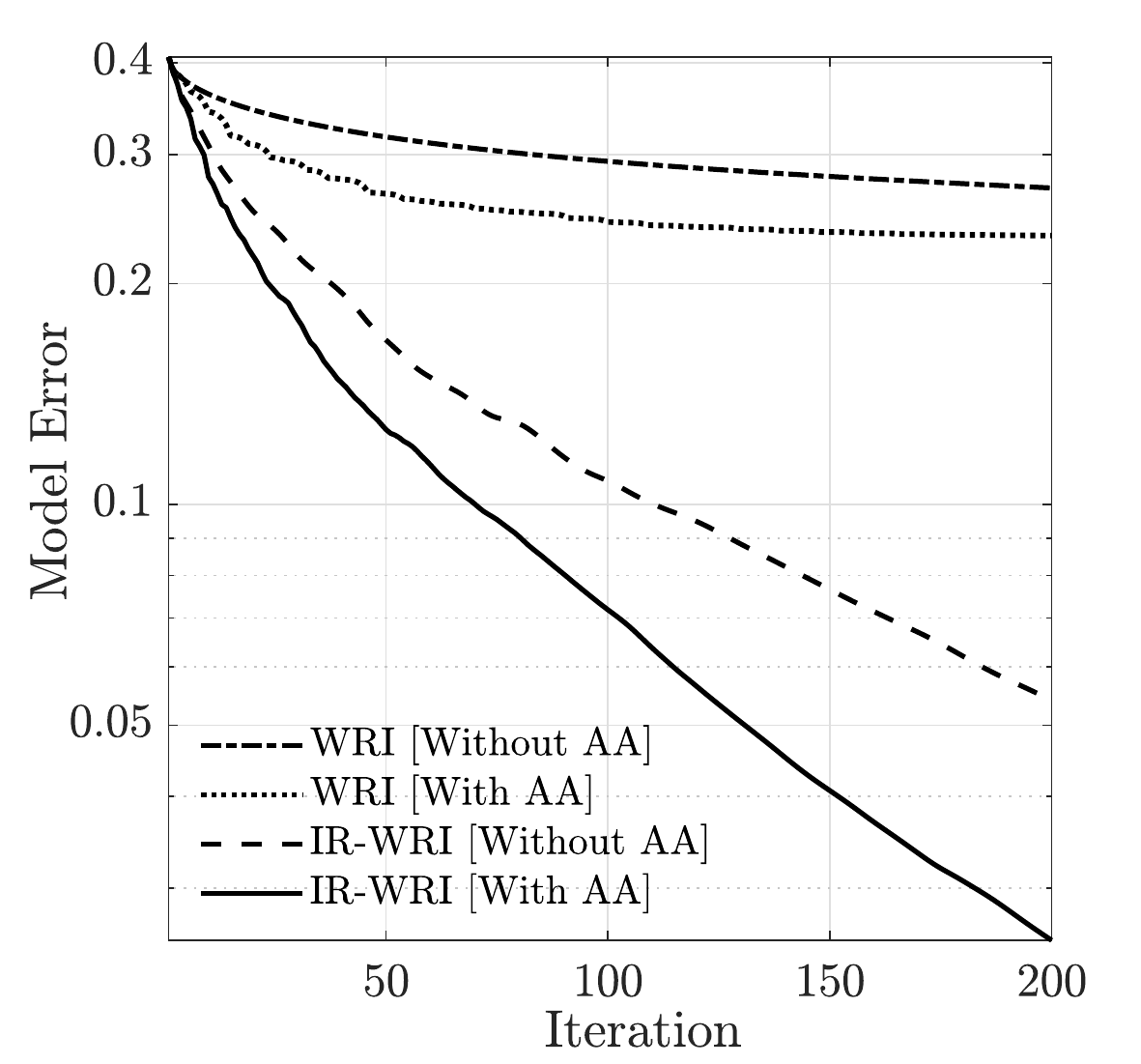}
   \caption{The checkerboard test. Model error curves versus iteration number for inverted models in Figure \ref{checker:blue}. 
   }
   \label{checker_MSE:blue}
 \end{figure}

The first example investigates the performance of the proposed accelerated strategy against a checkerboard model for both WRI \citep{van2013mitigating} and IR-WRI methods with BC. The model is composed of  a 1400~m$\times$1400~m homogeneous background model of velocity 1.5~km/s to which is added a  checkerboard perturbation model of velocity 2.5~km/s (Figure \ref{checker:blue}a). The acquisition setup consists of four sources located at the corners of the model and 276 receivers spaced 20~m apart along the four edges of the model. A Ricker wavelet with a central frequency of 10~Hz is used as a source term. We performed simultaneous inversion of frequencies 2.5 and 5~Hz using the background model as initial model. For the AA case, the size of the history is set to 10. Also, a fixed penalty parameter $\lambda=\alpha/\beta=10^{3}$ is used for all tests. The estimated models after 200 iterations obtained by WRI  and IR-WRI without AA are shown in Figure \ref{checker:blue}b and \ref{checker:blue}c. Then, we performed inversion using AA and the corresponding results are presented in Figure \ref{checker:blue}d (for WRI) and \ref{checker:blue}e (for IR-WRI). 
In the case of WRI with AA (Figure \ref{checker:blue}d), we see a remarkable improvement in the update of the model in comparison with WRI without AA (Figure \ref{checker:blue}b). Concerning IR-WRI, the extracted model with AA is closer to the true model, especially at the center (Figure \ref{checker:blue}e).  The corresponding model error curves versus iteration shown in Figure \ref{checker_MSE:blue} highlight two features. First, classical IR-WRI (dashed curve) clearly outperforms classical WRI (dash-dotted curve). Second, the convergence rates of the accelerated WRI (dotted curve) and IR-WRI (black curve) improve compared to those obtained with their classical implementation, which is consistent with the estimated models shown in Figure \ref{checker:blue}.
For the rest examples we test the performance of AA only with IR-WRI. 
 %
 %
 %
 %
\subsection{The Marmousi II model}

\begin{figure}[!htb]
   \centering
   \includegraphics[width=.45\textwidth]{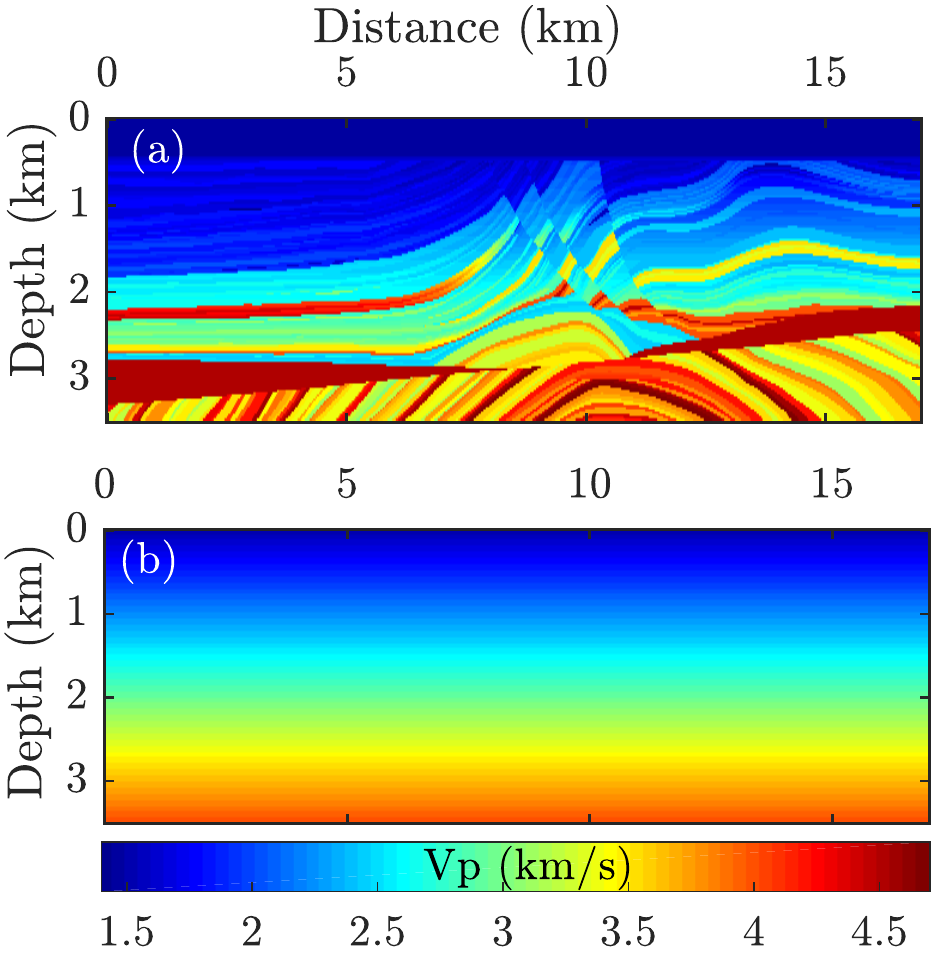}
   \caption{The Marmousi II test. (a) true velocity model and (b) initial velocity model.}
   \label{marm_true:blue}
 \end{figure}

For the second example, we consider the Marmousi II velocity model of dimension 3.5~km $\times$ 17~km (Figure \ref{marm_true:blue}a).
The model is re-sampled with a 25 m grid interval in both $x$ and $z$ directions. We consider a surface acquisition with 114 sources and 681 receivers spaced 150~m and 25~m apart, respectively. We start the inversion from a 1D velocity model linearly increasing from 1.5~km/s to 4.5~km/s (Figure \ref{marm_true:blue}b). Also, a Ricker wavelet with a dominant frequency of 10 Hz is used as the source function. For all the following experiments the size of the history of AA is set to $h=8$ experimentally to get a balance between performance and computational efficiency.  
We perform the inversion using both clean and noisy data under the same conditions. 
The frequencies range involved in the inversion is [3-15]~Hz with a frequency interval of 0.5~Hz. Mono-frequency batches were successively inverted when ten iterations per frequency batch is considered as stopping criterion of iteration. Besides, we used a fixed penalty parameter of $\lambda=\alpha/\beta=10^{6}$ during the iterations. Inversion is performed through two frequency paths, in which the updated model of the first path is set as the initial model for the second path.
%

\begin{figure}[!htb]
 	\centering
 	\includegraphics[width=0.8\textwidth]{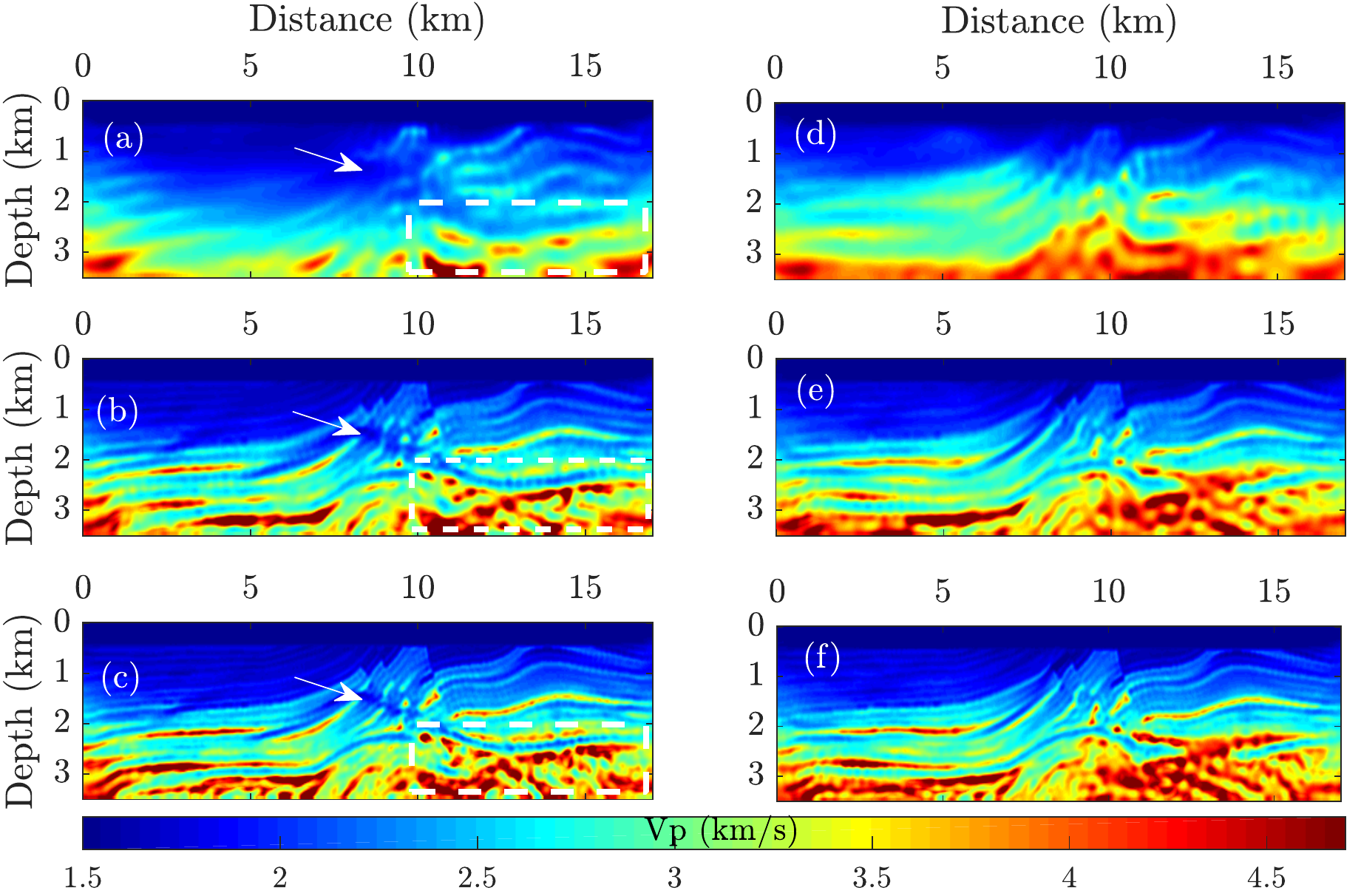}
 	\caption{The Marmousi II test. Inversion result of 3~Hz (first row), 5~Hz (second row) and 7~Hz (third row) frequency for IR-WRI (a-c) without AA and (d-f) with AA. In the left panel, some regions in the model are assigned by arrow and rectangle that demonstrate imperfection of conventional IR-WRI in comparison with its accelerated version (right column).} 
 	\label{marm_V3-5-7:blue}
\end{figure}

We first show some improvement achieved by IR-WRI at intermediate iterations when AA is implemented. 
Figure \ref{marm_V3-5-7:blue} shows the inversion results obtained by IR-WRI without and with  AA after 10, 50 and 90  iterations.
%
 
In the updated model by IR-WRI, we can recognize the failure of the method in capturing the kinematic components of the model (arrow and rectangle in Figure \ref{marm_V3-5-7:blue}a-c). 
In contrast, AA improves the reconstruct of these structures even at early iterations (Figure \ref{marm_V3-5-7:blue}d-f).
This observations show that AA can be considered as a  preconditioner of IR-WRI.

\begin{figure}[!ht]
	\centering
	\includegraphics[width=1\textwidth]{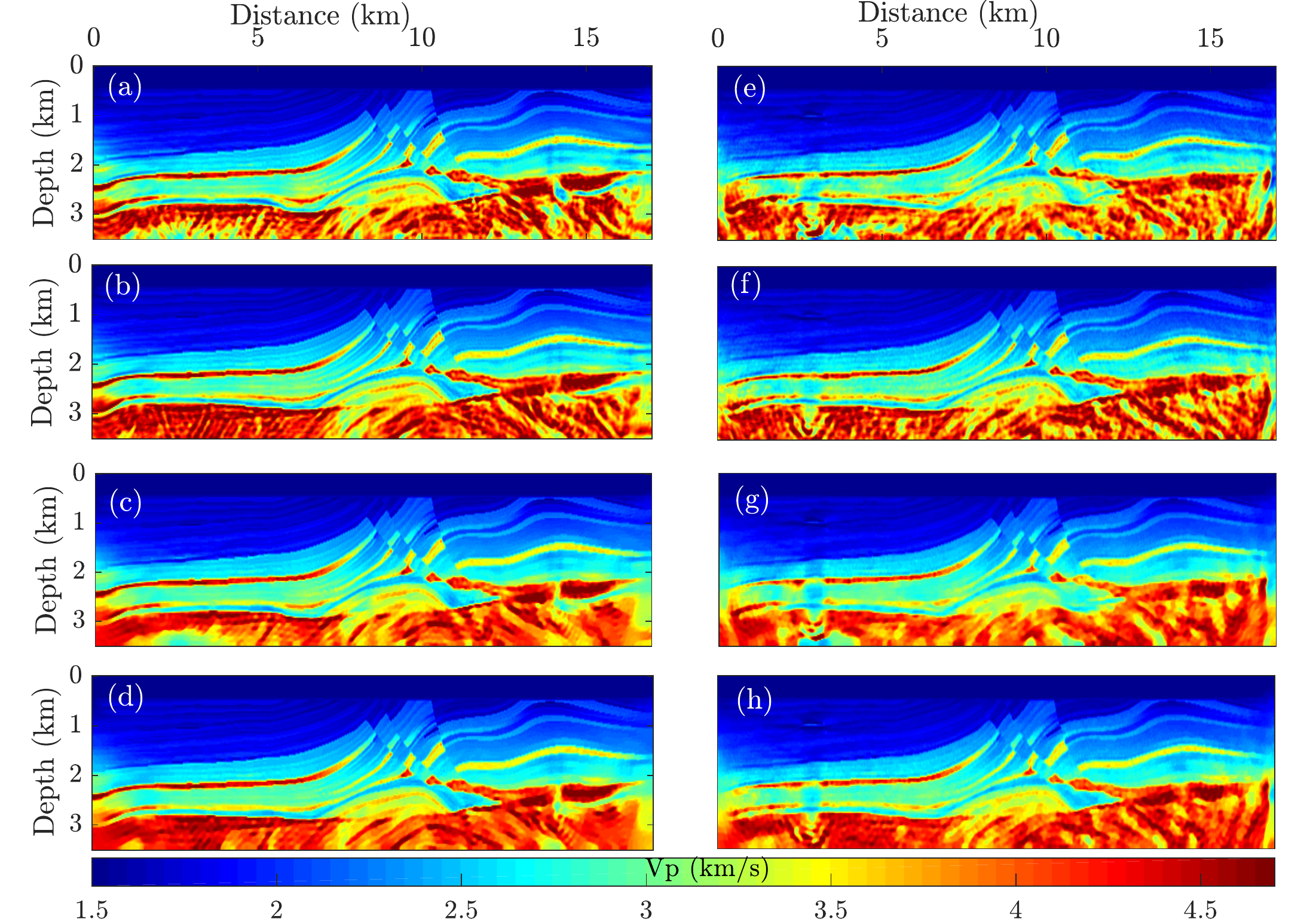}
	\caption{The Marmousi II test. The final inversion results after 340 iterations for (a) IR-WRI without AA, (b) IR-WRI with AA, (c) IR-WRI + TV, and (d) IR-WRI + TV + AA. (e-h) the same as (a-d) but for noisy data (S/N=5 dB).}
	\label{marm_Vfinal:blue}
\end{figure}

The final inversion results obtained from the Marmousi data are shown in Figure \ref{marm_Vfinal:blue}. IR-WRI without AA (Figure \ref{marm_Vfinal:blue}a) and with AA (Figure \ref{marm_Vfinal:blue}b) converge to similar results down to depth of 3~km. Note that, during the second inversion path, classical IR-WRI was able to reconstruct the selected regions in Figure \ref{marm_V3-5-7:blue}, where it was not able to reconstruct during the first path. However, for the deeper parts of the model, the accelerated version of IR-WRI outperformed the original version. 

\begin{figure}[!htb]
	\centering
	\includegraphics[width=.8\linewidth ,height=.8\linewidth]{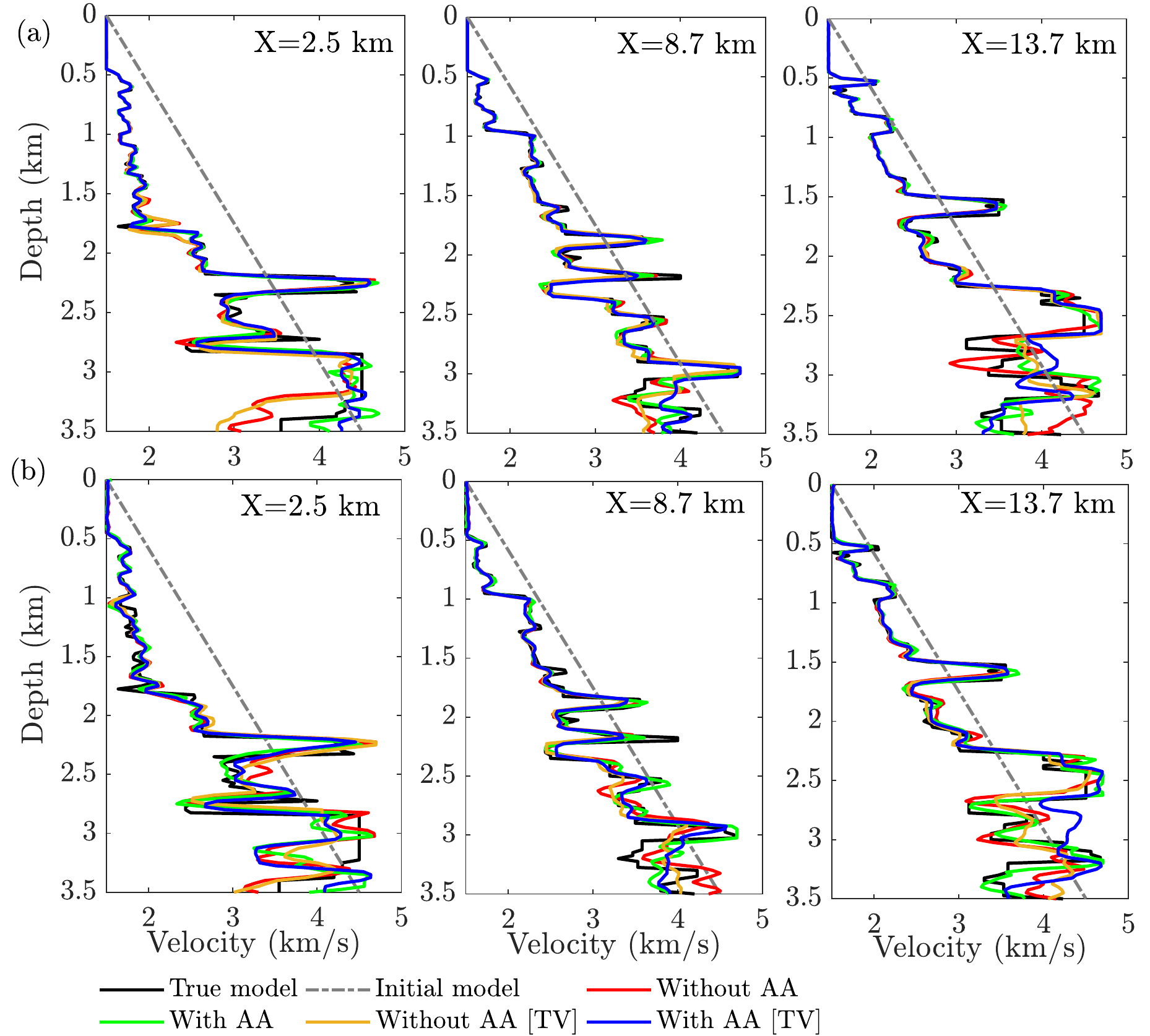}
	\caption{The Marmousi II test. 
	Top row: direct comparison between true model (black), IR-WRI (red).  IR-WRI + AA (green), IR-WRI + TV (orange), and IR-WRI + TV + AA (blue) at different locations specified by X in each panel. Bottom row: the same as top row but for noisy data. }
	\label{marm_velLog1:blue}
\end{figure}

\begin{figure}[!htb]
	\centering
	\includegraphics[width=0.8\textwidth]{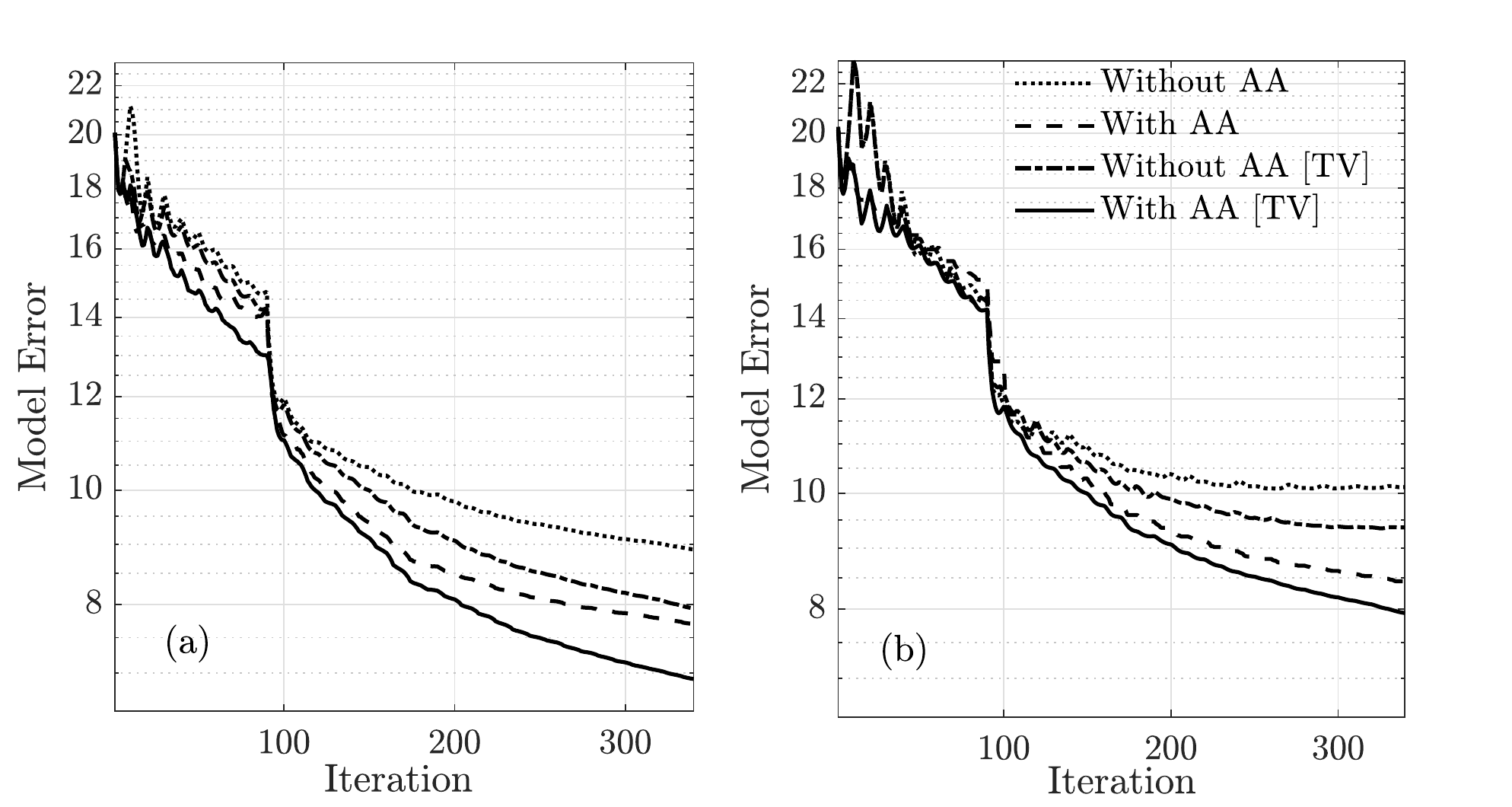}
	\caption{The Marmousi II test. Evolution of the model error versus iteration for different models in Figure \ref{marm_Vfinal:blue} for (a) noise free data and (b) noisy data.}
	\label{MSE_1:blue}
\end{figure}

Now we add the TV regularization to the inversion algorithm to see the performance of the AA with TV-regularized IR-WRI. Note that, in this case, the auxiliary primal-dual variables of the TV regularization, which are generated during variable splitting, also undergo the AA.
We set the penalty parameters according to \citet{aghamiry2019implementing}. 
Figures \ref{marm_Vfinal:blue}c-d show the final inversion results of TV regularized IR-WRI without and with AA. 
We observe that implementation of TV (Figure \ref{marm_Vfinal:blue}c) improves the quality of the result when comparing with the result of bounded IR-WRI (Figure \ref{marm_Vfinal:blue}a).
The result is further improved when we include the AA  (Figure \ref{marm_Vfinal:blue}d).
These observations can be seen more directly from the extracted vertical velocity profiles shown in Figure \ref{marm_velLog1:blue}a and from evolution of the model error curves displayed in Figure \ref{MSE_1:blue}a.

\begin{figure}[!htb]
	\centering
 	\includegraphics[width=.8\textwidth]{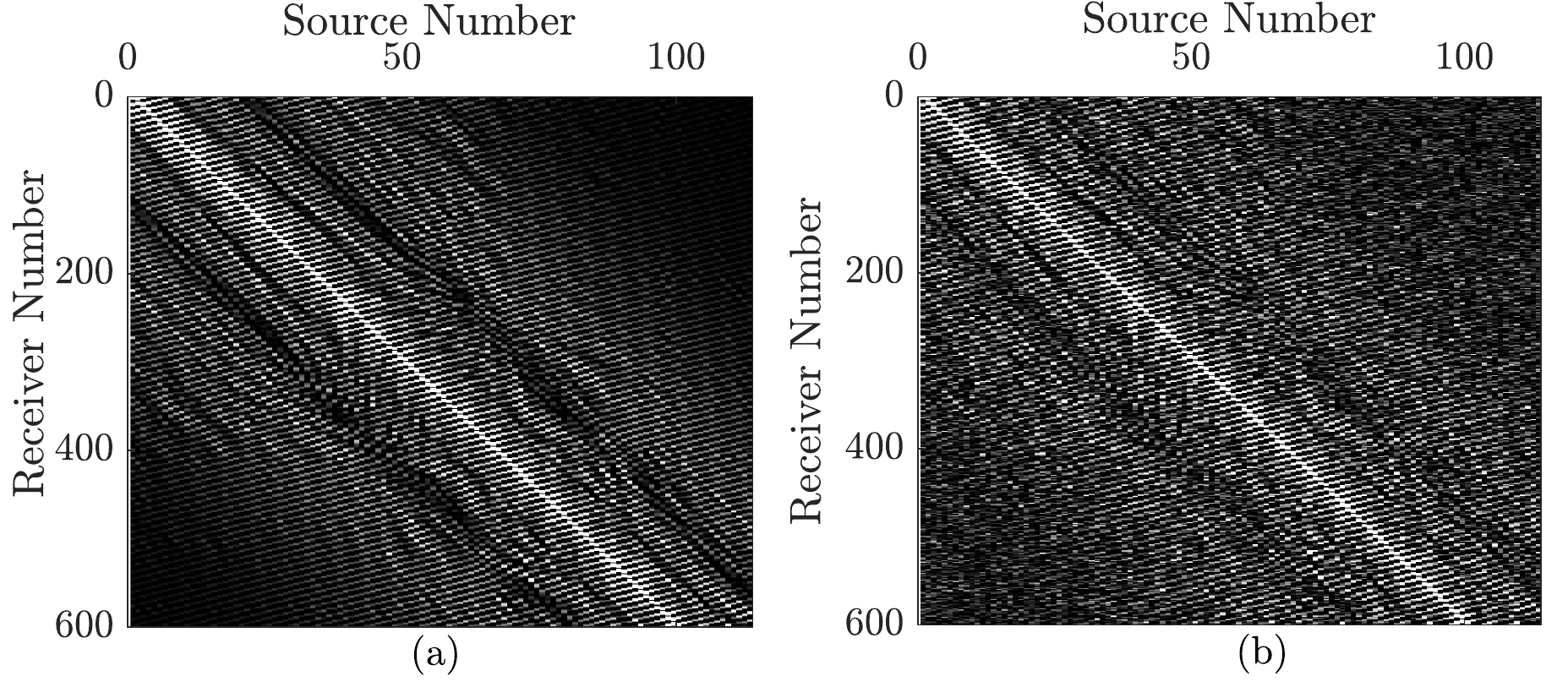}
	\caption{The Marmousi II test. Frequency-domain comparison between (a) noise-free and (b) noisy data (with S/N=5 dB) in source-receiver domain for frequency of 8~Hz. }
	\label{data_comp:blue}
\end{figure}

Also, we repeat the numerical tests for noise contaminated data. For this, low-pass filtered random noise of  signal-to-noise ratio (S/N) 5~dB is added to the data. 
In order to illustrate the noise strength, we show a frequency-domain comparison between the real part of noise-free and noisy data in the source-receiver coordinate system for the frequency of 8~Hz in Figure \ref{data_comp:blue}. The inversion framework remains the same as that used for the case of clean data. However, as investigated by \citet{aghamiry2019implementing}, the regularization parameter was increased a little to prevent the overfitting of data during the IR-WRI iterations. 
The final inversion results for IR-WRI, IR-WRI with AA, IR-WRI with TV, and IR-WRI with TV and AA are shown in Figures \ref{marm_Vfinal:blue}e-h, respectively. 
The associated vertical velocity profiles and evolution of the model error curves are also displayed in Figures \ref{marm_velLog1:blue}b and \ref{MSE_1:blue}b.
It can be seen that, for all inversion cases performed (with and without TV regularization and with and without noise), AA improves the convergence rate of IR-WRI. 

\subsubsection{On the effect of safeguarding and AA history}
\begin{figure}[!htb]
	\centering
	\includegraphics[width=.8\textwidth]{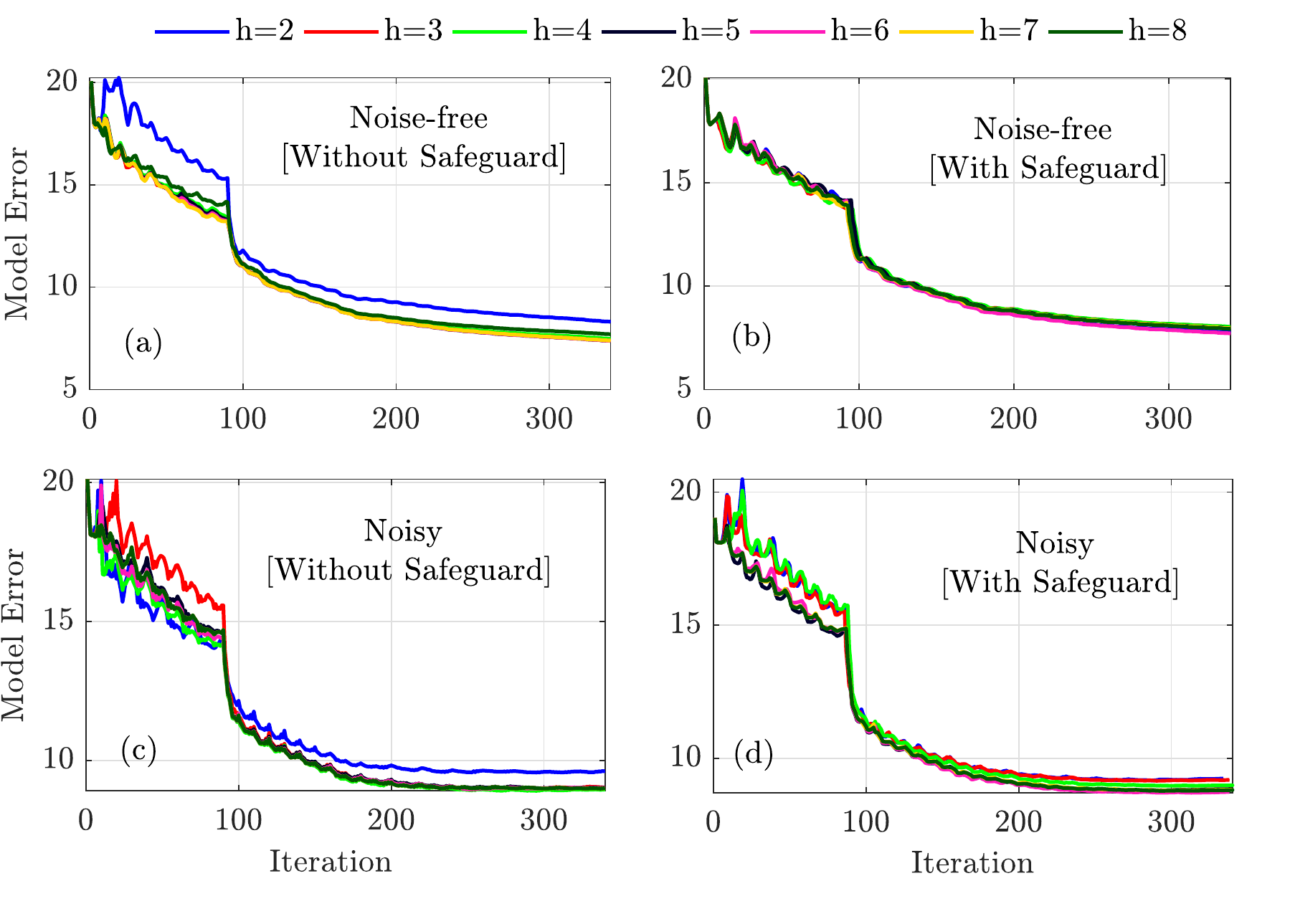}
	\caption{The Marmousi II test. The influence of history parameter on the evolution of model error for noise-free (top row) and noisy data (bottom row). (a,c) Results without safeguarding. (b,d) Same as (a,c) without safeguarding.}
	\label{MSE_marm_safe:blue}
\end{figure}

A pre-defined parameter for applying AA is the size of the history ($h$). This value may affect both regularization and safeguarding in addition to the performance of the AA algorithm. Usually, very small value of $h$ may not accelerate the inversion properly and may have a negative effect \citep{walker2011anderson}.  Conversely, a larger value requires more memory storage and may increase the ill-conditioning of the problem. Thus, there is a trade-off in the determination of $h$. We performed AA with seven values of $h= 2, 3, 4, 5, 6, 7, 8$ for the Marmousi II test with and without the safeguarding step for the case of noise-free and noisy data sets. The model error curves for these tests are shown in Figure \ref{MSE_marm_safe:blue} in which the top and bottom rows respectively correspond to noise-free and noisy data. Also, the left and right columns show the results of AA without and with safeguarding step. Regarding the effect of lower history value, we observe the weakness of AA for $h=2$ in noise-free case (Figure \ref{MSE_marm_safe:blue}). However, by applying the safeguarding step, the uncertainty of AA due to the choice of history decrease dramatically  (Figure \ref{MSE_marm_safe:blue}b). Regarding noisy data (Figures  \ref{MSE_marm_safe:blue}c-\ref{MSE_marm_safe:blue}d) we observe similar behaviour. 
%
%
\subsection{The 2004 BP model}
\begin{figure}[!htb]
   \centering
   \includegraphics[width=0.8\textwidth]{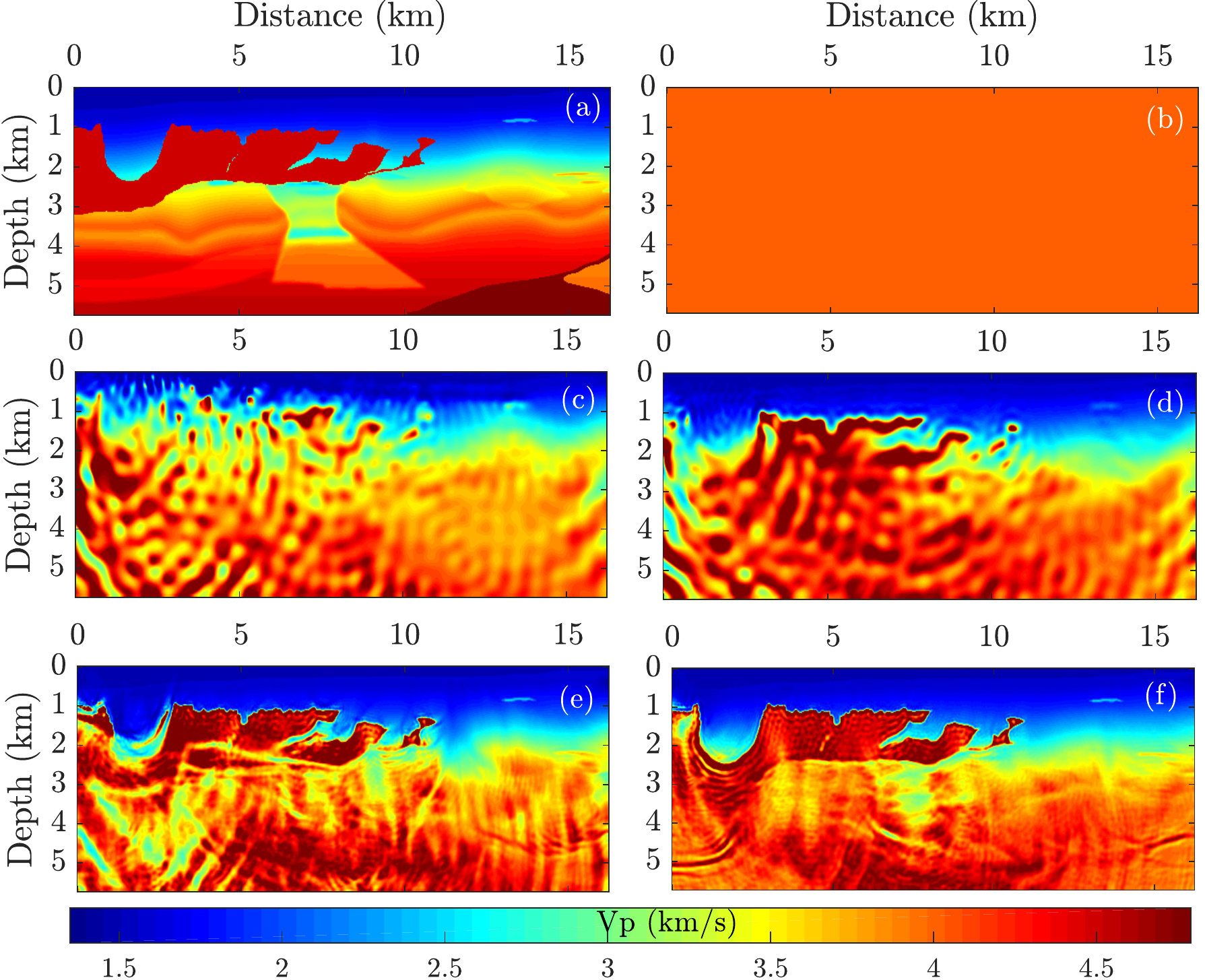}
   \caption{The BP salt model test. (a) true model and (b) initial model. (c-d) the inversion result of the first frequency batch (3-3.5 Hz)  without AA (c) and with AA (d). (e-f) the final inversion result obtained without AA (e)  and with AA (f).}
   \label{BP_v:blue}
\end{figure}

In the third example, we assess the performance of the proposed method against the left part of the challenging 2004 BP salt model \citep{billette20052004}. Here we use the resampled (with a grid interval of 25 m) and rescaled version of the model. The model covers a 5.725 km $\times$ 16.225 km area (Figure \ref{BP_v:blue}a). We consider a fixed-spread surface acquisition with 109 sources and 325 receivers uniformly distributed on the surface when a 10 Hz Ricker wavelet is used as the source signature.
The starting velocity model is a homogeneous model of 4 km/s (Figure \ref{BP_v:blue}b). We divide the inversion path into three frequency paths, [3-3.5]~Hz, [3-6]~Hz, [3.5-13]~Hz. The final estimated model of each path is used as the initial model for the next one. Inside each path, monofrequency inversion is performed with a 0.5 Hz frequency interval. We utilize $h=6$ to perform acceleration. The stopping criterion for iteration is a predefined maximum number of iteration or a predefined source misfit level. We set the maximum number of iterations equal to 35 for the first path and 20 for the rest.  

%

We do not show the results of TV regularization for this test because we wanted to see that even without regularization the AA helps IR-WRI greatly in building this challenging model with a crude starting model.
Definitely, implementing the TV regularization will improve the results as we observe for the Marmousi test.
Figures \ref{BP_v:blue}c,d show the inversion results of the first path (3-3.5 Hz) without AA (Figure \ref{BP_v:blue}c) and with AA (Figure \ref{BP_v:blue}d).
We can observe that traditional IR-WRI fails to recover the low-frequency information of the model properly. 
But the AA helps the IR-WRI to construct the top salt with correct kinematic informations.  

\begin{figure}[!htb]
   \centering
   \includegraphics[width=0.8\textwidth]{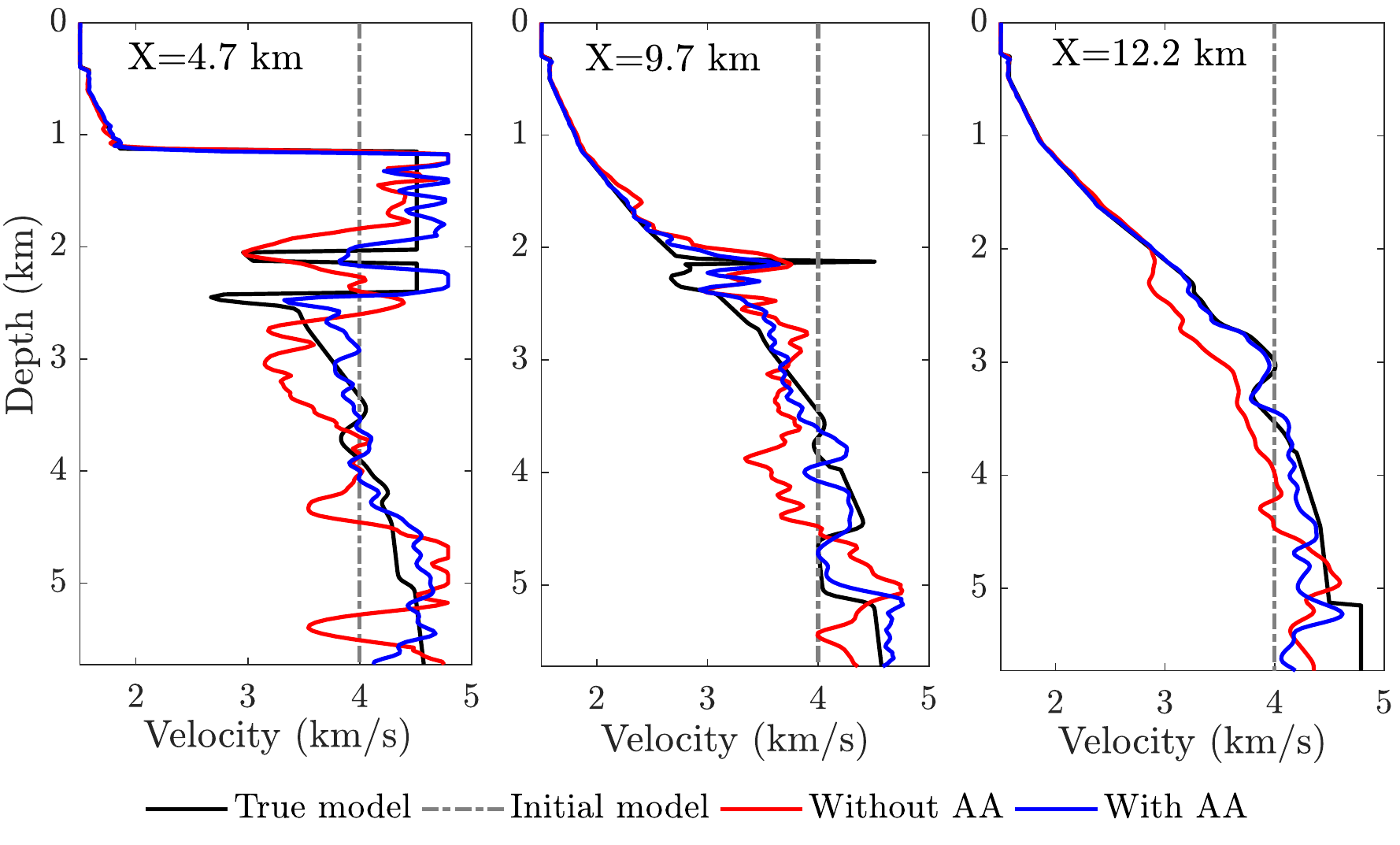}
   \caption{The BP salt model test. Direct comparison between true model (black), initial model (dashed), and the final results obtained by IR-WRI without AA (red) and with AA (blue) at different locations (specified by X).}
   \label{BP_logs:blue}
 \end{figure}

\begin{figure}[!htb]
   \centering
   \includegraphics[width=.4\textwidth]{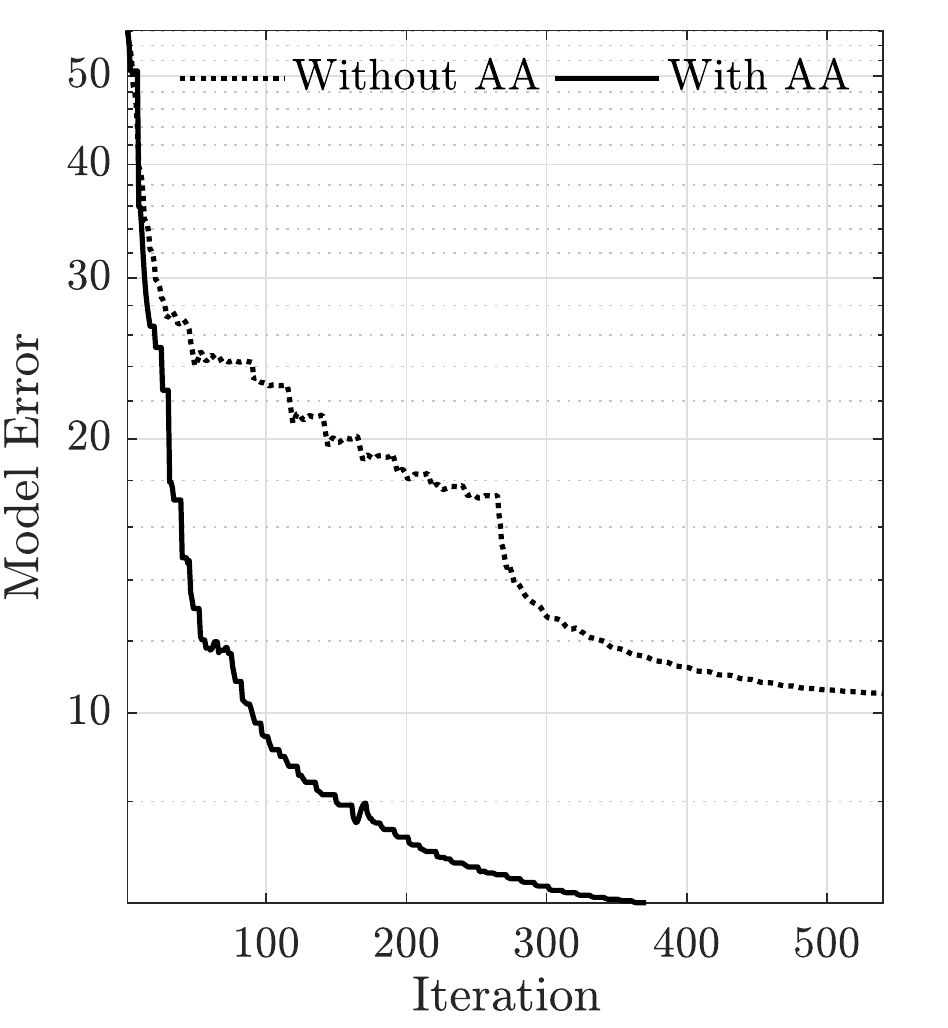}
   \caption{The BP salt model test. Evolution of the model error versus iteration number for IR-WRI without and with AA.}
   \label{BP_hist_err:blue}
 \end{figure}

We continue the inversion for the next paths. The final inverted velocity models are shown in Figure \ref{BP_v:blue}e (after 580 iterations without AA) and Figure \ref{BP_v:blue}f (after 371 iterations with AA). 
Estimated velocity models undergoes direct comparison in Figure \ref{BP_logs:blue}. We can observe that AA improves the estimate by properly following the structure of the true model (black).
This improvement can also be seen from the evolution of the errors as shown in Figure \ref{BP_hist_err:blue}.  
%
 
%

 \section{Discussion}
Despite the popularity of the AA in other fields of science, to the best of our knowledge, the study conducted by \citet{yang2020anderson} is the only application of the AA on seismic inversion. The experiments in  \citet{yang2020anderson} indicate the superiority of the AA over limited memory-BFGS (L-BFGS).  However,  as analyzed by \citet{fang2009two}, there is no generality in the advantage of AA over other Broyden's methods. The AA is known as the "off-the-shelf" acceleration method \citep{henderson2019damped}, i.e. which does not require "step-length" calculation as in the case of L-BFGS. The only requirement is to recast the original problem as a fixed point iteration. 
 
Our analysis in the framework of IR-WRI reveals some interesting features of AA, which distinguish our study from that of \citet{yang2020anderson}. First, we extend the application of the AA method to ADMM iterations by recasting the estimation of the primal-dual variables as a fixed-point iteration. Moreover, our methodology goes one step further by including useful prior information and regularization. We show that the auxiliary and dual variables of BTV regularization, which are created for handling non-differentiable functions based on splitting schemes, can be processed as extra fixed-point parameters. This generalization makes our algorithm to be more flexible and robust than the studies conducted by \citet{yang2020anderson} for FWI problem or \citet{zhang2019accelerating} for ADMM application in geometry optimization. Besides, our algorithm can consider two other options depending on the problem at hand. The first one is the regularization of the quadratic problem in the AA algorithm (equation \ref{eq:main}), i.e., known as damped-AA, which is already studied by \citet{zhang2018globally} and \citet{henderson2019damped}. The second one is the safeguarding step. The experiments for noise-free and noisy data show that applying this step improves the AA results and its robustness against of noise. As such, the IR-WRI with AA can be efficiently performed with small values of the history. 

We evaluate the proposed AA-based IR-WRI against three synthetic models. For the checkerboard model, we also apply the AA in the WRI method \citep{van2013mitigating}. The results show that with AA, the WRI performance is improved. However, the performance of IR-WRI without and with AA outperforms the WRI, which shows the importance of considering the dual variables as fixed-point parameters. Also, we show that IR-WRI with AA keeps its decreasing pattern in model error just like other reported experiments in other fields of science \citep{walker2011anderson} or time-domain seismic inversion in \citet{yang2020anderson}. 
For the Marmousi II and BP 2004 experiments, the scenario is somehow different. Although the general problem is to find the best-estimated model parameter, however, for each frequency batch (or even frequency path) we solve a new problem since after a few iterations, the frequency to be inverted is changed, and we reset the dual variables, $\mathbf{b}_{k}$, $\mathbf{d}_{k}$. As such, the AA history related to $\mathbf{b}_{k}$, $\mathbf{d}_{k}$ also resets for each frequency batch. In such cases, frequency-domain IR-WRI with AA can be seen as an AA with a periodic restart.  Therefore, one may not expect a regular decreasing pattern similar to the checkerboard test. Nevertheless still, IR-WRI with AA outperforms the conventional one. For example, the analysis of both noise-free and noisy data show that utilizing the AA scheme in the IR-WRI algorithm improves its performance, i.e. the calculated model errors and source residual curves of IR-WRI without AA require more iterations to obtain such accuracy yielded by the IR-WRI with AA. 

\section{CONCLUSIONS} 
We recast the IR-WRI iteration as a general fixed-point iteration to improve the convergence speed of IR-WRI with Anderson acceleration (AA). The accelerated IR-WRI keeps a pre-defined history of the previous iterations and builds the new iteration by a linear combination of the history. The combination weights are determined at each iteration for the optimal convergence by solving a least-squares problem. 
We analyzed the performance of the proposed acceleration scheme through numerical examples using a simple checkerboard test and the Marmousi II and 2004 BP benchmark models. The results show that a small history (less than 10) is enough to have a good performance. Also, they show that damping and safeguarding could improve the performance of IR-WRI with AA.  
Future work will concentrate on investigating the proposed method for the case of multiparameter IR-WRI for elastic physics. 

\section*{ACKNOWLEDGMENTS}  
This study was partially funded by the WIND consortium (\textit{https://www.geoazur.fr/WIND}), sponsored by Chevron, Shell and Total. The authors are grateful to the OPAL infrastructure from 
Observatoire de la Côte d'Azur (CRIMSON) for providing resources and support. This work was granted access to the HPC resources of IDRIS under the allocation A0050410596 made by GENCI.
\bibliographystyle{seg}
\newcommand{\SortNoop}[1]{}

\end{document}